\documentclass[twocolumn,aps,pre,showpacs,floatfix,showkeys]{revtex4}

\usepackage{amsfonts}
\usepackage[english]{babel}
\usepackage{amsmath,amssymb}
\usepackage[all,cmtip]{xy}

\def\be{\begin{equation}}
\def\ee{\end{equation}}
\def\ba{\begin{eqnarray}}
\def\ea{\end{eqnarray}}

\newtheorem{definition}{Definition}

\newtheorem{theorem}{Theorem}

\def\beq{\begin{equation}}
\def\eeq{\end{equation}}
\def\ba{\begin{eqnarray}}
\def\ea{\end{eqnarray}}
\begin{document}

\title[GROUP ENTROPIES]{Group entropies, correlation laws and zeta functions}


\author{Piergiulio Tempesta}
\address{Departamento de F\'{\i}sica Te\'{o}rica II, Facultad de F\'{\i}sicas, Ciudad Universitaria, Universidad
Complutense, 28040 -- Madrid, Spain}
\email{p.tempesta@fis.ucm.es}


\begin{abstract}
The notion of group entropy is proposed. It enables to unify and generalize many different definitions of entropy known in the literature, as those of Boltzmann--Gibbs,
Tsallis, Abe and Kaniadakis. Other new entropic functionals are presented, related to nontrivial correlation laws characterizing universality classes of systems out of
equilibrium, when the dynamics is weakly chaotic. The associated thermostatistics are discussed.  The mathematical structure underlying our construction is that of formal group theory,
which provides the general structure of the correlations among particles and dictates the associated entropic functionals.
As an example of application, the role of group entropies in information theory is illustrated and generalizations of the Kullback--Leibler divergence are proposed.
A new connection between statistical mechanics and zeta functions is established. In particular, Tsallis entropy is related to the classical Riemann zeta function.
\end{abstract}

\volumeyear{year}
\volumenumber{number}
\issuenumber{number}

\maketitle


\section{Introduction}
In many physical contexts, as well as in economy, biology and social sciences, the nature of the \textit{correlations} among the parts or the subsystems constituting
a given system is crucial for the understanding of the underlying dynamics. A possible example is offered by the present international economical context. Indeed, the simplifying hypothesis of microeconomical independence in local or national markets appear to be inadequate to describe the evolution of the present economical scenario: the strong and unexpected correlations that the crisis has generated among previously unrelated economic and financial entities make inadequate the foresight of models based on this hypothesis (for a general perspective see e.g. the monograph \cite{MS}). Many experimental observations in condensed matter physics, nuclear physics, astrophysics as well as in several fields of social sciences reveal the need for a general formulation of statistical mechanics that might proceed from the kind of correlation experimentally observed, and deduce coherently the associated thermostatistics.

Nowadays, the celebrated Boltzmann--Gibbs (BG) statistical mechanics is recognized as the appropriate theory to describe the thermodynamics of a very large class of physical systems, ubiquitous in nature, at thermal equilibrium. A paradigmatic example is the case of systems with short--range interactions and short--time memories. Also, BG statistical mechanics has been successfully adopted in the description of critical phenomena and in nonequilibrium statistical mechanics, when the ergodic or the chaotic hypothesis are assumed \cite{Gal1}--\cite{Huang}.


However, there are frequent situations when at least some of the basic assumptions of the BG theory are violated (for instance when ergodicity is not assumed).
In \cite{Tsallis1}, the approach called nonextensive statistical mechanics was introduced to extend the applicability of the standard theory to these "pathological", but still common situations.

In most of the cases when a weak chaotic regime is observed, e.g. in several systems exhibiting long--range interactions, scale invariance \cite{TGS}, multifractal structure, etc., and recently, in hadronic physics \cite{CMS}, the nonextensive approach offers an adequate theoretical framework for the comprehension of the thermodynamics of the underlying dynamics \cite{Tbook}. A huge literature exists on this theory and its multiple applications in science (for a regularly updated bibliography, see \cite{TEMUCO}). However, it should be noticed that the kind of \textit{correlation} between different subsystems contemplated in this approach, although very relevant for the applications, is just one of the many possible types we can hypothesize.

In this work, we address the following question: there exists a unified theoretical framework, suitable for constructing generalized entropies and related thermostatistics for a wide class of systems \textit{out of equilibrium}, correlated in a nontrivial way?  We provide an affirmative answer, by extending the nonextensive approach in order to englobe a very general class of correlation laws among subsystems of a given system. Under suitable hypotheses, to each choice of the correlation law, it corresponds an entropic functional and a related microscopical description of the dynamics. From this point of view, \textit{it is the correlation that dictates the thermostatistics and the correct entropy to be used, and not viceversa}: the entropy is \textit{not} postulated, but its functional form emerges naturally from the class of interactions we wish to consider.

Our theory reposes on the notion of \textit{universal formal group}. Formal groups have been proposed by Bochner \cite{Boch}, with the aim of generalizing Lie groups and algebras. In the last decades, the theory of formal groups has been widely investigated for its crucial role in many branches of pure mathematics \cite{Haze}, \cite{Serre1}. It is especially relevant in algebraic topology (cobordism theory \cite{BMN}, theory of genera \cite{H}, homology theory \cite{Ray}) and in the theory of elliptic curves \cite{Serre2}. In \cite{Tempesta1}--\cite{Tempesta3}, formal groups have also been used  to construct a generalization of the Bernoulli polynomials and of the Riemann zeta function.

We will show that each realization of the Lazard universal formal group enables us to define a possible \textit{universality class} of statistical systems, in terms of the correlation law satisfied by the random variables associated to the considered system, and describing its observables. Each correlation law in turn defines a specific quantum calculus. Also, for each class we can introduce in a natural way an appropriate entropy of trace--form type, constructed by using the finite operator calculus \cite{Rota}, and analyze the corresponding generalized nonextensive statistical mechanics. Important physical constraints, like concavity or Lesche stability, are also considered, since they are indispensable in order to have physically satisfactory entropies.

It should be noticed that the meaning of "universality" in our context presents some similarities with the usual one, as referred to critical phenomena. In any case, we will not be concerned with critical phenomena in this work.

The physical need for such family of new entropies is evident, for instance, from the physical example of a system of N interacting over-damped particles, recently published \cite{Andrade}. The entropy for this system, which is neither Boltzmann's nor Tsallis'one, falls in the class of universal group entropy and can be easily studied in our formalism.

However, the group entropies are not exclusively designed for physical purposes. Indeed, due to the ubiquitous role of the notion of entropy in modern science \cite{Wehrl}, the possible applications of the proposed construction are manifold. For instance, another non trivial application of our entropic functionals emerges in the context of Information Theory. We will show that these entropies provide a class of information measures, including the Shannon information measure as a particular case. Also, we propose a generalization of the Kullback--Leibler divergence, that allows us to define a new set of tests measuring the difference between two given probability distributions.

A further result is the connection we establish between statistical mechanics and number theory. This fascinating topic has an intriguing history, dating back to the 70's with the works of Montgomery and Odlyzko, relating the Gaussian unitary ensemble with the zeros of the Riemann zeta function $\zeta(s)$ \cite{frontiers}. In \cite{julia}, it has been shown that the Riemann zeta function is the canonical partition function of a free bosonic gas. Also, recently the zeros of $\zeta(s)$ have been related to Landau levels \cite{ST} for a charged particle on a planar surface in an electric potential and uniform magnetic field. A quantum mechanical model whose spectrum is the sequence of prime numbers has been proposed in \cite{mussardo}.

Our construction is quite different from other number--theoretical approaches to statistical mechanics. Under appropriate hypotheses, with a class of universality we can associate a zeta function, constructed by using the same realization of the Lazard group used to define the corresponding entropy. The family of zeta functions considered here have been introduced in \cite{Tempesta1} and further studied in \cite{Tempesta3}. The first nontrivial case is the Tsallis class, which corresponds to the celebrated Riemann zeta function.

We mention that other generalizations of nonextensive statistical mechanics are known in the literature, based on different kinds of deformations of logarithmic and exponential functions (see, for instance, \cite{Naudts}, \cite{KLS}, Chapter VI of \cite{Tbook} and references therein). The present one also differs from the superstatistics scenario proposed in \cite{Beck}. Our approach aims to unravel both the group and number--theoretical content of the notion of entropy, and seems new. Also, it is constructive: all the entropic functionals are obtained in a explicit form.

Work is in progress on a quantum version of the theory developed here \cite{Tempesta}. In particular, group entropies emerge as natural measures in the Hilbert space of the states of a multipartite system as entanglement measures  \cite{Horodecki}.

The paper is organized as follows. In Section 2, we discuss a family of logarithm--type functions, obtained in the context of finite difference calculus, along with the formulation of G. C. Rota. In Section 3, the notion of group entropy is introduced, based on the previous construction. In Section 4, the universality classes related to group entropies are constructed; their thermodynamic properties are studied in Section 5.
In Section 6, we develop a measure--theoretic setting for interpreting the group entropies as information measures.
In Section 7, the group entropies are associated with zeta functions. Some open problems are discussed in the final Section 8.

\section{Difference operators and logarithmic functions}

In order to construct generalized thermostatistics, and the related entropic functionals, we start introducing a family of logarithmic functions, obtained from suitable representations of certain finite--difference operators. For the sake of clarity and to offer a self--contained exposition, some basic concepts are reviewed. The mathematical apparatus is kept to a minimum.

Let $\mathcal H$ a space of sufficiently regular functions of a real variable $x$; a possible choice is, for instance, a Banach algebra of functions. Let us denote by $T$ the shift operator, whose action on a function $f \in \mathcal{H}$ is given by $T f\left(x\right)=f\left(x+\sigma\right)$. Here $\sigma$ is a real parameter, whose absolute value can be interpreted as the width of a regular, equally spaced lattice of points $\mathcal{L}$. We essentially adopt the formalism of \cite{Rota}.

In order to define the class of entropic functionals of interest for this work, we will consider operators expressed as finite Laurent series in shift operators \cite{LTW1}:
\begin{equation}
\Delta_{r}=\frac{1}{\sigma}\sum_{n=l}^{m}k_{n}T^{n}\text{,}\quad l\text{,
}m\in\mathbb{Z}\text{,}\mathbb{\quad}l<m\text{,\quad}m-l=r\text{,}\label{2.9}
\end{equation}
where $\{k_{n}\}$ are real constants such that
\begin{equation}
\sum_{n=l}^{m}k_{n}=0, \quad\sum_{n=l}^{m} nk_{n}=c. \label{2.10}
\end{equation}
and $k_m\neq0$, $k_l\neq0$. We choose $c=1$, to reproduce the derivative $D$ in the continuum limit, when the
lattice spacing $\sigma$ goes to zero.

\begin{definition}
\textit{A difference operator of the
form (\ref{2.9}), which satisfies equations (\ref{2.10}), is said to be a delta operator of
order $r$, if it approximates the continuous derivative up to terms of order $\sigma^r$.}
\end{definition}

As eq. (\ref{2.9}) involves $m-l+1\,$constants
$k_{n}$, subject to just the two conditions (\ref{2.10}), we can fix all constants
$k_{n}$ by choosing $m-l-1$ further conditions. A possible choice is, for instance, to set
\beq
\sum_{n=l}^{m}|n|^{\ell}k_{n}=K_\ell, \qquad \ell=2,3,...,m-l. \label{constraints}
\eeq
with $K_{\ell}$ suitable real numbers.

The main idea underlying our construction is to represent delta operators in logarithmic form, in terms of a suitable function.

\begin{definition}
\textit{We call logarithmic representation of the delta operator (\ref{2.9}) the correspondence $T\leftrightarrow x^{\sigma}$, that defines an isomorphism $\mathcal{I}$ between the space of shift--invariant operators and the space of functions $f\in\mathcal{H}$.}
\end{definition}

The main definition of the Section is the following.
\begin{definition}
\textit{We call generalized logarithm the function
\begin{eqnarray}
\noindent \nonumber Log_{G}(x)=\frac{1}{\sigma}\sum_{n=l}^{m}k_{n}x^{\sigma n}, \quad l, m\in\mathbb{Z},\\
 \mathbb{\quad}l<m,\quad m-l=r, \quad x>0 \label{Log}
\end{eqnarray}
with the constraints (\ref{2.10})--(\ref{constraints}), i.e. the image of the operator (\ref{2.9}) under the isomorphism $\mathcal{I}$.}
\end{definition}

The following Lemma motivates the choice of the name "logarithm" for the function (\ref{Log}).

\noindent\textbf{Lemma 1.}
\textit{The following property holds:
\beq
\lim_{\sigma \rightarrow 0} Log_{G}(x)= \ln x,
\eeq
under the conditions (\ref{2.10})--(\ref{constraints}).}

\textbf{Proof}.
The constraints (\ref{2.10})--(\ref{constraints}) ensure that, in the limit $\sigma \rightarrow 0$, the discrete derivatives of the family (\ref{2.9}) tend to the continuous derivative $\partial_x$. If we put $x=e^t$, it implies that the function $\frac{1}{\sigma}\sum_{n=l}^{m}k_{n}x^{\sigma n}\equiv \frac{1}{\sigma}\sum_{n=l}^{m}k_{n}e^{\sigma n t}$ tends to $t= \ln x$ when $\sigma \rightarrow 0$.

\par
Notice also that the first of the two conditions (\ref{2.10}) implies that
\beq
Log_{G} (1) = 0. \label{Log1}
\eeq
Several examples of interesting logarithmic functions can be obtained by using the logarithmic representation. For instance, if we use the entropic parameter $q$, via the identification $\sigma\equiv 1-q$, the Tsallis logarithm \cite{Tsallis1}, \cite{Tbook} corresponds to the discrete derivative of order $r=1$ $\Delta^{+}=\frac{T-1}{\sigma}$:
\beq
Log_{\mathcal{T}}(x)=\frac{x^{1-q}-1}{1-q}. \label{tsallis}
\eeq

\noindent The Kaniadakis logarithm \cite{Kaniad1} is the indicator of the operator of order two $\Delta_{s}=\frac
{T-T^{-1}}{2\sigma}$ :
\beq
Log_{{K}}(x)=\frac{x^{\kappa}-x^{-\kappa}}{2 \kappa}. \label{kaniad}
\eeq
(we put here $\sigma = \kappa$ in accordance with the standard notation).

\section{Group Entropies}

Motivated by the previous construction, we propose one of the central notions of this work.

\noindent\textbf{Definition 4.}
\textit{Consider a discrete probability distribution $\{p_i\}_{i=1,\cdots,W}$, normalized as
\beq
\sum_{i=1}^{W}p_i=1.\label{norm}
\eeq
We call group entropy the functional
\beq
S_{G}(p):=k \sum_{i=1}^{W}p_i Log_{G}(\frac{1}{p_i}), \label{entropy}
\eeq
where $Log_{G}$ denotes the generalized logarithm (\ref{Log}) with the constraints (\ref{2.10}), (\ref{constraints}), and $k\in \mathbb{R^{+}}$.}

In physical contexts, we can typically identify $k$ with the Boltzmann constant $k_B$; otherwise,  as in information theory, we tacitly assume $k=1$. The reason for the denomination of \textit{group entropy} for $S_G$ comes from its connection with the \textit{universal formal group}: it will determine the corresponding correlations for the class of physical systems under examination.

As will be explained below, the freedom in the determination of the set of constants $k_n$, left by the conditions (\ref{2.10}), (\ref{constraints}) can be used to impose the requirement of concavity, Lesche stability \cite{Lesche}, etc. for our group entropies.

The entropies previously defined belong to the class of \textit{trace form entropies}. This class is very general, but does not include other functional forms, like Renyi's entropy \cite{Renyi}, also interesting in several applications. The classical Boltzmann--Gibbs entropy
\beq
S_{\mathcal{B}}=-k\sum_{i=1}^{W}p_i \ln p_i \label{BG}
\eeq
is obtained from (\ref{entropy}) in the limit $\sigma\rightarrow 0$. The Tsallis and the Kaniadakis one correspond to the choices (\ref{tsallis}), (\ref{kaniad}).
Let us now construct some new examples. Consider for instance the difference operators
\begin{eqnarray}
\nonumber &   \Delta_{III}=\frac{T-2T^{-1}+T^{-2}}{\sigma}, \quad \Delta_{IV}=\frac{T^2-\frac{3}{2}T+\frac{3}{2}T^{-1}-T^{-2}}{\sigma},  \\
\nonumber &  \Delta_{V}=\frac{T^3-2 T^{2}+2 T -2T^{-1}+T^{-2}}{-\sigma},
\end{eqnarray}

and so on. The corresponding logarithms are
\begin{eqnarray}
\nonumber\noindent Log_{G_{III}}(x)&=&\frac{1}{\sigma}\left(x^{\sigma}-2 x^{- \sigma}+x^{-2\sigma}\right), \\
\nonumber\noindent Log_{G_{IV}}(x)&=&\frac{1}{\sigma}\left(x^{2\sigma}-\frac{3}{2}x^{\sigma}+\frac{3}{2}x^{-\sigma}-x^{-2\sigma}\right), \\
\nonumber\noindent Log_{G_{V}}(x)&=&\frac{1}{\sigma}\left(x^{3\sigma}-2x^{2\sigma}+2x^{\sigma}-2x^{-\sigma}+x^{-2\sigma}\right).
\end{eqnarray}
Consequently, we introduce the entropies
\begin{eqnarray}
&& S_{G_{III}}(p):=\frac{k}{\sigma}\sum_{i=1}^{N}p_{i}\left( p_{i}^{2 \sigma}-2p_{i}^{\sigma}+p_{i}^{-\sigma}\right), \label{SIII} \\
&& S_{G_{IV}}(p):=\frac{k}{\sigma}\sum_{i=1}^{N}p_{i}\left( p_{i}^{-2 \sigma}-\frac{3}{2} p_{i}^{-\sigma}+\frac{3}{2}p_{i}^{\sigma}-p_{i}^{2\sigma}\right), \label{SIV} \\
&& \nonumber S_{G_{V}}(p):=\frac{k}{\sigma}\sum_{i=1}^{N}p_{i}\left( p_{i}^{-3 \sigma}-2 p_{i}^{-2\sigma}+2 p_{i}^{-\sigma}-2 p_{i}^{\sigma}+p_{i}^{2\sigma}\right),
\\ \label{SV}
\end{eqnarray}

\noindent and so on.

\noindent Here the roman sub--indices are used in order to distinguish the logarithms and the associated entropies according to the order of the discrete operator they come from. The entropic forms (\ref{SIII})--(\ref{SV}) at the best of our knowledge are new. It would be desirable to produce an axiomatic formulation of the notion of group entropy, along the lines of Shannon--Khinchin--Abe's approach \cite{shannon}--\cite{Shannon2}. Observe that group entropies can be easily constructed in order to fulfill the first three of the classical Khinchin
axioms \cite{Khinchin}. Indeed, for a suitable choice of the set of coefficients $\{k_n\}$ and at least in a suitable
range of values of $\sigma$, the general entropy (\ref{entropy}) 1) is continuous;
2) reaches its maximum for the equiprobability distribution $p_i=1/W, i=1,\ldots, W$; 3) satisfies the condition $S_{G}(p_1,p_2,\ldots,p_W,0)=S_{G}(p_1,p_2,\ldots, p_W)$,
which amounts to say that the addition of an event of zero probability does not affect the value of the entropy.
It should be noticed that, although the entropy (\ref{entropy}) is certainly continuous in the case of a finite number of microstates $W$, in general is no longer continuous for $W=\infty$, when is not continuous the Boltzmann--Gibbs entropy either \cite{HT}. The fourth Khinchin axiom, regarding additivity, obviously does not hold in the present nonextensive context. Precisely, in our approach, the formula (\ref{nonadditive}) below,
concerning the "composability" of a system, \textit{is no longer an axiom}, but a property, coming naturally from our group--theoretical construction.

Along these lines, in the recent work \cite{Hanel}, a classification of trace--form entropies has been recently proposed. By
relaxing the fourth Shannon--Kinchin axiom,  universality classes of entropies are introduced and an explicit expression of them in terms of the incomplete Gamma function (Gamma entropies) is provided.

Notice that there is a simple alternative way to derive the group entropies defined above. Indeed, the functions (\ref{entropy}) are characterized by the following interesting generating formula:
\beq
-k\left[\Delta(\alpha)\sum_{i=1}^{W}p_i^{\alpha}\right]_{\alpha=1}=S_{G}. \label{deltaentropy}
\eeq
Here $\Delta(\alpha)$ stands for the operator (\ref{2.9}), applied to the variable $\alpha$. Formula (\ref{deltaentropy}) specializes in
\beq
-k\left[\frac{d}{d\alpha}\sum_{i=1}^{W}p_i^{\alpha}\right]_{\alpha=1}=S_{\mathcal{B}}
\eeq
Relation (\ref{deltaentropy}) is inspired by a similar one, due to Abe \cite{Abe1} for Tsallis' entropy, involving the Jackson derivative of the quantity $\sum_{i=1}^{W}p_i^{\alpha}$.

In the next section, the intimate connection between the proposed entropies and group theory will be clarified. In particular, the correlation laws underlying the proposed group entropies will be derived.

\section{Universality classes and correlation laws}

\subsection{Main definitions}

We wish to propose a classification of statistical systems in terms of universality classes related to group entropies.

Given a commutative ring $R$ with identity, and the ring $R\left\{ x_{1},\text{ }%
x_{2},\ldots\right\} $ of formal power series in the variables $x_{1}$, $x_{2},\ldots,$ with coefficients in $R$, a commutative one--dimensional formal group
over $R$ is a formal power series $\Phi \left( x,y\right) \in R\left\{
x,y\right\} $ such that \cite{Boch}
\begin{eqnarray}
\nonumber &   1)\qquad \Phi \left( x,0\right) =\Phi \left( 0,x\right) =x \\
\nonumber &  2)\qquad \Phi \left( \Phi \left( x,y\right) ,z\right) =\Phi \left( x,\Phi
\left( y,z\right) \right) \text{.}
\end{eqnarray}
When $\Phi \left( x,y\right) =\Phi \left( y,x\right) $, the formal group is
said to be commutative. The existence of an inverse formal series $\varphi
\left( x\right) $ $\in R\left\{ x\right\} $ such that $\Phi \left( x,\varphi
\left( x\right) \right) =0$ follows from the previous definition.

Perhaps the most general definition of trace form entropy comes from the theory of formal groups.

\textbf{Definition 5.}
\textit{Consider the formal power series over the polynomial
ring $\mathbb{Q[}c_{1}, c_{2},...]$ defined by
\begin{equation}
G\left( t\right) = \sum_{i=0}^{\infty} c_i \frac{t^{i+1}}{i+1},
\label{I.1}
\end{equation}
\noindent with $c_0=1$, called \textit{formal group exponential}. Let $F\equiv G^{-1}$ be the compositional inverse of (\ref{I.1}):
\begin{equation}
F\left( s\right) =\sum_{i=0}^{\infty} \gamma_i \frac{s^{i+1}}{i+1} \label{I.2}
\end{equation}
so that $G\left(F\left( t\right) \right) =t$. We have $\gamma_{0}=1, \gamma_{1}=-c_1, \gamma_2= \frac{3}{2} c_1^2 -c_2,\ldots$.
The Lazard universal formal group law \cite{Haze} is defined by the formal power series
\begin{equation}
\Phi \left( s_{1},s_{2}\right) =G\left( F \left(
s_{1}\right) +F\left( s_{2}\right) \right). \label{FGL}
\end{equation}}


Let us analyze the structure of the correlation law among subsystems for the family of entropies (\ref{entropy}), i.e. their "composability" property. Assume that $\mathcal{S}$ is an abstract statistical system (physical, biological, etc.), composed by two independent subsystems $A\subset\mathcal{S}$  and $B\subset\mathcal{S}$. A general property of the group entropy (\ref{entropy}) of $\mathcal{S}$ is nonadditivity. Precisely, the following result holds.

\begin{theorem}
\textit{The group entropy (\ref{entropy}) satisfies the following nonadditive property:
\beq
S_G(A+B)=\Phi(S_G(A),S_G(B)) \label{nonadditive}
\eeq}
\end{theorem}

\textbf{Proof}.
It it easy to prove that eq. (\ref{nonadditive}) is true for the group entropy $S_G$ if and only if it holds for the group logarithm $Log_G(x)$ as well. If we use the exponential representation $x\leftrightarrow e^t$, we obtain that $Log_G(x) \equiv G(t)$. Since $t=G^{-1}(s):=F(S)$, the relation $G(t_1+t_2)=G(F(s_1)+F(s_2))$ holds, which is the thesis.

As a natural consequence of the previous result, a huge class of deformed algebraic structures and calculi closely connected both with the classical one (i.e. the algebra of real numbers) and the standard q--calculus can be derived from formal group laws.

\noindent\textbf{Definition 6.}
\textit{The generalized sum of two numbers x and y is given by the universal formal group law (\ref{FGL})
\beq
x\oplus_{\gamma} y=\Phi\left(x,y\right).\label{correlation}
\eeq
Here $\gamma$ is the set of parameters appearing in (\ref{I.2}).}

\textbf{Remark 1.} The Definition 4 of Group Entropy can be easily generalized to the case when the group exponential is indeed not simply a function of $e^t$ (as in Tsallis' case and the other ones considered above), but a generic \textit{formal series} of the type (\ref{I.1}). In this case we will talk about the \textit{universal group entropy}. In the following, a physical example when this more general situation is contemplated is discussed.

We propose here a possibly new definition of \textit{universality classes} for statistical systems. Essentially, it allows to identify systems sharing the same correlation law among their subsystems.

\textbf{Definition 7.}
\textit{A universality class of statistical systems is a set of systems that satisfy the following properties.
\begin{itemize}
\item The correlation law between two independent subsystems of a given system is expressed by the formal group law (\ref{FGL}).
\item The thermostatistics associated is governed by the entropy constructed starting from the corresponding group logarithm and exponential.
\end{itemize}}

As a matter of fact, once $F$ and $G$ are known, the entropy associated can be easily deduced, as shown in the subsequent discussion.


\subsection{Some relevant universality classes}

The simplest example is the \textit{Boltzmann--Gibbs universality class}.
The entropic functional for this class is the Boltzmann--Gibbs entropy (\ref{BG}). The composition law (\ref{correlation}) is defined by the choice $G(t)=t$, and the same for $F$. We get:
\beq
x\oplus_{\gamma} y=x+y.
\eeq
i.e. the additive formal group law.

The \textit{Tsallis universality class} corresponds to the choice
\beq
G_{\mathcal{T}}(t)=\frac{e^{\sigma t}-1}{\sigma} \label{GT}
\eeq
for the group exponential, with the inverse
\beq
\qquad F_{\mathcal{T}}(s)=\frac{1}{\sigma}\ln(1+\sigma s).
\eeq
The related $q$--calculus is defined by
\beq
x\oplus_{q} y=x+y+(1-q)xy,
\eeq
(with $ \sigma=1-q$), which is the multiplicative formal group law.

For the \textit{Kaniadakis universality class}, we obtain
\beq
G_{\mathcal{K}}(t)=\frac{e^{\sigma t}-e^{-\sigma t}}{2\sigma} \label{GT}
\eeq
for the group exponential. Its inverse is
\beq
\qquad F_{\mathcal{K}}(s)=\frac{1}{\sigma}\ln(s\sigma +\sqrt{s^2 \sigma^{2}+1}).
\eeq
We have (by putting $ \sigma=k$)
\beq
x\oplus_{k} y=x\sqrt{1+k^2y^2}+y\sqrt{1+k^2x^2},
\eeq
which represents a specific realization of the \textit{Euler formal group law for elliptic integrals}. Besides, as is well known, it coincides with the composition law of relativistic momenta in special relativity.

An example of entropy belonging to the more general class, defined in Remark 1, when the group logarithm is not of the form (\ref{Log}) but possesses
a general expansion of the type (\ref{I.2}), is
Abe's entropy. It is related to the following logarithmic functional (\cite{Abe1})
\beq
Log_{\mathcal{A}}(x)=\frac{x^{\left(\sigma-1\right)}-x^{\left(\sigma^{-1}-1\right)}}{\sigma-\sigma^{-1}}.
\eeq

By using the previous considerations, we can easily generalize the previous logarithm by introducing the two--parametric functional
\beq
Log_{a,b}(x)=\frac{x^{a}-x^{b}}{a-b}, \label{temp}
\eeq
which reproduces Abe's one for $a=\sigma-1, b=\sigma^{-1}-1$. It has been already considered in the Borges--Roditi construction of an entropic functional \cite{BR}. In this case, the related group exponential, obtained by using the isomorphism $\mathcal{I}$, reads
\beq
G_{\mathcal{A}}(t)=\frac{e^{at}-e^{bt}}{a-b}\label{Abel}.
\eeq
The formal group corresponding to (\ref{Abel}), giving the interaction rule for this class, is known in the literature as the \textit{Abel formal group}, defined by \cite{BK}
\beq
\Phi_{\mathcal{A}}(x,y)=x+y+\beta_1 xy+ \sum_{j>i} \beta_i\left(xy^{i}-x^{i}y\right). \label{AFG}
\eeq
The coefficients $\beta_n$ in (\ref{AFG}) can be expressed as polynomials in $a$ and $b$ (see Proposition 3.1 of \cite{BK}):
\beq
\beta_n=\frac{(-1)^{n-1}}{n!(n-1)}\prod_{\overset{i+j=n-1}{i,j\geq 0}} (ia+jb).
\eeq


Concerning the algebraic structure of the theory, observe that we can introduce a new multiplication law, inspired by the previous construction.

\noindent\textbf{Definition 8.}
\textit{Given two real numbers x and y, the product $x\otimes y$ is defined by the relation
\beq
Log_{G}(x \otimes y) = Log_{G} x + Log_{G} y.\label{mult}
\eeq}

%

\noindent We recover easily the known cases.

\textit{i)} For the case of the Boltzmann class, $x \otimes y = xy$, i.e. the multiplication (\ref{mult}) reduces to the standard pointwise multiplications of real numbers.

\textit{ii)} For the Tsallis class, we have \cite{Borges}
\beq
x \otimes_q y= \left[x^{1-q}+y^{1-q}+1\right]^{\frac{1}{1-q}}
\eeq

\textit{iii}) For the Kaniadakis class, we obtain \cite{KLS}
\beq
x \otimes_k y= \frac{1}{k} \sinh \left[\frac{1}{k} arcsinh \left(kx \right)arcsinh\left(xy\right)\right]
\eeq

However, with these composition laws we do not get a priori standard algebraic structures: as has been already noticed for the specific example of the Tsallis class, there is no guarantee that the distributivity property be satisfied.

\subsection{An interesting physical example}

Very recently, in \cite{Andrade} a system of $N$ over--damped interacting particles moving in a narrow channel has been studied. The equations for the velocities of the particles are
\beq
\mu \overrightarrow{v_i}=\sum_{j\neq i} \overrightarrow{J}\left( \overrightarrow{r_i} -\overrightarrow{r_j}\right)+\overrightarrow{F^{e}}(\overrightarrow{r}_i)+\eta(\overrightarrow{r_i},t)
\eeq
where $\overrightarrow{v_i}$ is the velocity of the $i$th particle, $\overrightarrow{J}\left( \overrightarrow{r_i} -\overrightarrow{r_j}\right)$ is a short--range repulsive particle--particle interaction, $\mu$ is the effective viscosity of the medium, $\overrightarrow{F^{e}}(\overrightarrow{r}_i)$ is an external force, $\eta(\overrightarrow{r_i},t)$ is a thermal noise with zero mean and variance $<\eta^2>=k_{B}T/\mu$. One of the main results of \cite{Andrade} is that, for intermediate temperatures of the thermal bath in which the system is immersed, the entropy of the system is given by a linear combination of Tsallis entropy for $q=2$ and the Boltzmann--Gibbs entropy. The novel type of thermostatistics proposed by these authors reveals the need for a generalization of the entropic functionals commonly used to include new ones inspired by concrete applications. From this point of view, the notion of group entropy provides a simple and unifying approach to this issue. It is easy to see that the example \cite{Andrade} can be easily accomodated in the framework previously proposed. Indeed, the group exponential is
\beq
G(t)=at+b \frac{e^{(q-1) t}-1}{q-1}.
\eeq
Note that this case corresponds to the more general situation discussed in the Remark 1, and can be immediately generalized to any linear combination of known group entropies. We get
\beq
S_{G}(t)=k \sum_{i=1}^{W}\left[p_{i}\ln p_{i} +p_{i}(1-p_{i})\right].
\eeq

\section{The thermodynamical properties of the theory: Legendre structure and H theorem}
A natural question is what thermal statistics come from the generalized entropies discussed above. In this Section we address this problem from a formal point of view. We show that the \textit{Legendre structure of classical thermodynamics} for a system in a stationary state would be preserved by any of the extensions which can be obtained within our mathematical framework. Other issues, as possible generalizations of the zeroth law of thermodynamics \cite{MPP1} are not addressed. Essentially, here we shall discuss the maximization of group entropies under appropriate constraints: we advocate a generalized maximum entropy principle, following \cite{AT}, \cite{Kaniad2}.

First, consider the \textit{microcanonical ensemble} for an isolated system in a stationary state. In this case, the only constraint is
\beq
\sum_{i=1}^{W}p_i=1.\label{norm}
\eeq

As a consequence, the optimization of $S$ yields equal probabilities, i.e. $p_i=1/W, \forall i$. Therefore, we have
\beq
S_G=k Log_G W,
\eeq
which reduces to the celebrated Boltzmann formula $S_{BG}=k\ln W$ in the case of uncorrelated particles.
Now we consider a generalized \textit{canonical ensemble}, i.e. a system in thermal contact with a reservoir.  Assume that $p_i(\epsilon_i)$ is a normalized and monotonically decreasing distribution function of $\epsilon_i$. We can think of the numbers $\epsilon_i$ as the values of a physically relevant observable, for instance the value of the energy of the system in its $i$th state. We define the \textit{internal energy} U in a given state as
\beq
U=\sum_{i=1}^{W}\epsilon_i p_i(\epsilon_i) .
\eeq
Consider the variational problem of the existence of a stationary distribution $\widetilde{p}_i(\epsilon)$. We introduce the functional
\beq
L=S_{G}[p]-\alpha\left[\sum_{i}p(\epsilon_i)-1\right]-\beta\left[\sum_{i=1}^{W}\epsilon_i p_i(\epsilon_i))-U\right],
\eeq
where $\alpha$ and $\beta$ are Lagrange multipliers. By imposing the vanishing of the variational derivative of this functional with respect to the distribution $p_i$, we get the stationary solution
\beq
\widetilde{p}_i=\frac{E(-\alpha-\beta (\epsilon_i-\widetilde{U})}{Z},
\eeq
with $Z=\sum_{i=1}^{W}E(-\alpha-\beta (\epsilon_i-\widetilde{U})$, and $E(\cdot)$ is an invertible function. However, only in particular cases (e.g. for the Boltzmann, Tsallis, Kaniadakis,
Borges--Roditi entropies) this function can be identified with the inverse of $Log_{G}$, according to the analysis of the role of generalized exponentials in thermodynamics performed in \cite{Kaniad2}.
Nevertheless, for generalized entropies constructed as a realization of the universal formal group that no longer belong to the trace--form class, one can contemplate
further cases in which $E(\cdot)$ can be set to be the inverse of $Log_{G}$.

Notice that a closed expression for the inverse of $Log_{G}$ can be analytically determined in very specific situations, since the inversion of a formal group logarithm would involve the solution of
polynomial equations of high degree.  Although we will keep the discussion at a formal level, however, we can assume that at least a numerical interpolating solution, in several cases and for specific subsets of the space of parameters, be available.

As usual, the parameter $\alpha$ can be eliminated by means of the constraint $\sum_i p(\epsilon_i)\frac{\delta L}{\delta p(\epsilon_i)} \mid_{p=\widetilde{p}}=0$.
If we perform the Legendre transform of $Log_G(Z)$, we get the interesting relation
\beq
Log_{G}(Z)+\beta U=S_{G},
\eeq
from which we deduce immediately the relation
\beq
\frac{\partial S_G}{\partial U}=\frac{1}{T},
\eeq
with $T\equiv 1/\beta$. In this context, $T$ plays the role of a local temperature, a priori function of space and time. In a nonequilibrium stationary state, the usual definition of temperature, as given by the zeroth law does not apply.  The previous discussion also implies the generalized form of the free energy:
\beq
\mathcal{F}=U-TS_{G}.
\eeq

A natural question is whether the group entropy satisfies an analog of the H--theorem. As has been shown in \cite{Abe2}, it is possible to prove the validity of the H--theorem for a large class of entropic forms. Precisely, assume that the group logarithm (\ref{Log}) be an invertible function, and piecewise monotonic. Consider the master equation
\beq
\frac{dp_i}{dt}=\sum_{j=1}^{W}\left(A_{ij}p_j-A_{ji}p_i\right),
\eeq
where $A_{ij}$ denotes the transition probability per unit time from the state $j$ to the state $i$. If we assume the principle of microscopic reversibility \cite{Landsberg}, which implies the relation
\beq
A_{ij}=A_{ji},
\eeq
then, under the previous assumptions we obtain
\beq
\frac{dS[p]}{dt}\geq 0.
\eeq

In order for an entropic functional to satisfy a maximum entropy principle, it is sufficient to ascertain that it verifies the following \textit{concavity property}. Consider two sets $\{p_i\}$ and $\{p'_i\}$ of probabilities taken from the same set of $W$ possibilities. Let $\lambda$ be a real parameter, satisfying $0<\lambda<1$. According to \cite{Tsallis1}, define the intermediate probability law
\beq
p''_i=\lambda p_i + (1-\lambda)p'_i.
\eeq
A direct calculation shows that the entropies (\ref{SIII})--(\ref{SV}), in a suitable interval of values of $\sigma$, satisfy the condition
\beq
S(\{p''_{i}\}) \geq \lambda S(\{p_i\}) + (1-\lambda S(\{p'_i\}).
\eeq


Another crucial aspect, for thermodynamical applications, is the \textit{experimental robustness} of the group entropy. It means that, given two distributions $\{p_i\}_{i=1,\ldots,W}$ and $\{p'_i\}_{i=1,\ldots,W}$ whose values are slightly different,
\beq
\forall \epsilon \quad \exists \delta >0 \quad s. t. \quad \|p-p'\| < \delta \Longrightarrow \left| \frac{S[p]-S[p']}{S_{max}}\right|< \epsilon,
\eeq
where, given a vector $x$, $\|x\|_1$ denotes the $L_1$ norm $\sum_{i=1}^{W}x_i$, and $S_{max}$ denotes the maximum value of the entropy. This property, also called Lesche stability \cite{Lesche} and equivalent to uniform continuity, is ensured under the hypothesis that the group exponential be at least piecewise differentiable. We point out that this requirement is especially important for physical applications (for a recent discussion, see \cite{Abe3}); however it can be disregarded in other contexts, as, for instance, Information Theory.

%

\section{Group entropies as generalized Kullback--Leibler divergences}

Following the classical approach by Shannon \cite{shannon}, \cite{Shannon2}, we show here that group entropies can naturally be interpreted as generalized information measures. This entails a proper definition of the functionals of the class (\ref{entropy}) on a continuous setting. As an important application, we introduce a generalization of the Kullback--Leibler relative entropy. This provides a very large set of measures of divergence between two different probability distributions, or alternatively of tests allowing to discriminate between two different hypotheses. Here, as in \cite{DBM}, we propose a formulation of our approach in a measure--theoretical setting.

Precisely, let $(X, \Sigma, \mu)$ be a measure space, where as usual $X$ is a set, $\Sigma$ a $\sigma$--algebra over $X$, and $\mu: \Sigma\rightarrow \mathbb{R}^{+}$ a measure. A $\Sigma$--measurable function $p:X\rightarrow \mathbb{R}^{+}$ will be called a \textit{probability distribution function} (p.d.f.) if
\beq
\int_{X} p d \mu = 1.
\eeq
The \textit{probability measure} induced by a pdf $p$ is defined by
\beq
P(E)=\int_Ep(x)d \mu(x), \qquad \forall E \in \Sigma.
\eeq

In the following, we will assume that $(X, \Sigma, \mu)$ is a $\sigma$--finite measure space, and that our probability measures are absolutely continuous with respect to $\mu$. We will also identify functions differing on a $\mu$--null set only. As a consequence of the classical Radon--Nikodym theorem, we will discuss on the same footing information measures coming from p.d.f.'s and from probability measures, since we can identify them up to a $\mu$--null set, with $P<<\mu$.


We can now propose the main definition of this Section.

\noindent\textbf{Definition 9.}
\textit{Given a $\sigma$--finite measure space $(X, \Sigma, \mu)$, the group entropy of a pdf p is given by
\beq
S_{G}(p):= \int_{X}p(x) Log_{G}\frac{1}{p(x)}d \mu(x), \label{inf}
\eeq
or equivalently in terms of a probability measure P absolutely continuous w.r.t. $\mu$,
\beq
S_{G}(p):= \int_{X}Log_{G}\left(\frac{d P}{d \mu}\right)^{-1}d P, \label{inf2}
\eeq
\noindent provided the integral on the right exists.}

Here $\frac{d P}{d \mu}$ denotes the Radon--Nikodym derivative of the measure P with respect to $\mu$. Observe that we put $k=1$, as customarily in Information Theory. The Shannon entropy of the pdf $p$ is obtained immediately from the previous definition:
\beq
S(p)=-\int_X p(x)\ln p(x) d \mu.
\eeq

In a similar way, we can introduce a generalization of the standard Kullback--Leibler relative entropy (KL) \cite{KL}, that we call \textit{relative group entropy}. In information theory, given two different probability distributions, the KL entropy is a measure of the expected number of additional bits required to code samples from P if, instead using a code based on P, we use a code based on Q.

\noindent\textbf{Definition 10.}
\textit{Let P and Q be two probability measures, and Q be absolutely continuous with respect to P. The relative group entropy is defined as
\beq
D_{G}(P\parallel Q)=-\int_{X} Log_{G}\left(\frac{dQ}{dP}\right)^{-1} dP,
\eeq
or equivalently, if $\mu$ is a measure, as
\beq
D_{G}(p\parallel q)=-\int_{X} p(x) Log_{G}\left(\frac{q(x)}{p(x)}\right)^{-1} d\mu.
\eeq}

The previous definition implies the useful relation
\beq
S_{G}(P)=D_{G}(P\parallel \mu).
\eeq

Our definition includes as a special case that proposed by Tsallis in \cite{Tsallis KL}. Also, it differs from that of Yamano \cite{Yamano}.

Let us study some properties of the relative group entropy. First, consider a \textit{uniform} distribution on a compact support of length $W$. We deduce that
\beq
D_{G}(p,1/W)=Log_{G}(W)-W^{\sigma} S_{G}(p),
\eeq
i.e. the RGE essentially measures the departure of a given group entropic functional from its value at equiprobability.

If we impose that $x Log_{G}(x)$ is a convex function, the relative group entropies belong to the class of Csisz\'ar $f$--divergence measures \cite{Csiszar}. In this case, the main property of the relative group entropy is a generalization of Gibbs' inequality. For simplicity, we will state it in the discrete case.

\textbf{Theorem 3.}
\textit{Let $\{p_i\}$ and $\{q_i\}$ two sets of probabilities. Assume that $x Log_{G}(x)$ is a convex function. Then the relative group entropy satisfies the inequality
\beq
D_{G}(P\parallel Q)=\sum_{i=1}^{W} p_i Log_{G} \frac{p_i}{q_i}\geq 0, \label{Gibbs}
\eeq
and the equality holds if and only if $p_i=q_i$ for all $i=1,\ldots, W$.}

\textbf{Proof}.
It suffices to observe that, if the function $\varphi(x):=x Log_{G} (x)$ is a convex one, we can use the classical Jensen inequality \cite{Jensen}. Indeed, if we have a set of nonnegative real numbers $\{p_1,\cdots,p_W\}$ and $\{q_1,\cdots,q_W\}$, with  $\sum p_i= \mathcal{S}_1, \sum q_i= \mathcal{S}_2$, then
\ba
\nonumber && \sum_{i=1}^{W}p_i Log_{G} \frac{p_i}{q_i}=\mathcal{S}_2 \sum_{i=1}^{W} \frac{q_i}{\mathcal{S}_2}\varphi(\frac{p_i}{q_i})\geq\mathcal{S}_2\varphi\left(\frac{\sum_{i=1}^{W} p_i}{\mathcal{S}_2}\right)= \\
&& \mathcal{S}_1 Log_{G} \frac{\mathcal{S}_1}{\mathcal{S}_2}.
\ea
Now, if $\mathcal{S}_1=\mathcal{S}_2=1$, the inequality (\ref{Gibbs}) follows. Due to the condition $Log_{G}(1)=0$, when $p_i=q_i$ for all $i=1,\ldots, W$ we have $D_{G}(P\parallel Q)= 0$.

This result justifies the interpretation of relative group entropies, in a measure--theoretic setting, as criteria of information divergence. However, a priori we can also construct non Csizs\'ar --type distances, for instance by relaxing the above convexity condition or, for instance, by constructing generalized Chernoff $\alpha$--distances \cite{Chern}. Further study is in progress in this direction.

\section{Dirichlet zeta functions associated with group entropies}

In this section, we clarify the number--theoretic content of the previous theory. The main result can be stated as follows: provided some technical hypotheses are satisfied, \textit{there exists a Dirichlet zeta function associated with each universality class of a suitable type}. In order to make the argument transparent, let us consider first the simplest nontrivial case, which is that of the Tsallis universality class. Quite interestingly, it is associated with the classical \textit{Riemann zeta function}.

First, observe that Tsallis logarithm (7), under the exponential representation $x\leftrightarrow e^t$, corresponds to the group exponential $\frac{e^{(1-q) t}-1}{1-q}$.

Let $\Gamma \left( s\right) =\int_{0}^{\infty
}e^{-t}t^{s-1}dt$ be the Euler $\Gamma$--function. Thus, if \textit{Re} $s>1$, and $q<1$, we have
\beq
\frac{1}{\Gamma \left(
s\right) }\int_{0}^{\infty }\frac{1}{\frac{e^{(1-q) t}-1}{1-q}}t^{s-1}dt=\frac{1}{(1-q)^{s-1}}\zeta \left( s\right).
\eeq
\noindent More generally, motivated by the previous example and by the definition of group entropy, let us consider the reciprocal of the following formal group exponential
\begin{equation}
G(t)=\frac{1}{\sigma}\sum_{n=l}^{m}k_{n}e^{\sigma n t}, \quad l, m\in\mathbb{Z}, \mathbb{\quad}l<m,\quad m-l=r, \label{expLog}
\end{equation}
with the constraints (\ref{2.10})--(\ref{constraints}) and $\sigma>0$. We assume that $1/G(t)$ is a $C^{\infty }$ function over $\mathbb{R}_{+}$, rapidly
decreasing at infinity. Compute (whenever possible) the formal expansion
\begin{equation}
\frac{1}{G\left( t\right) }=\sigma\sum_{n=1}^{\infty }a_{n}e^{-n \sigma t}\text{.}
\end{equation}

\noindent Thus, the function
\begin{equation}
L\left( G,s\right) =\frac{1}{\Gamma \left( s\right) }\frac{1}{\sigma^{s-1}}\int_{0}^{\infty }\frac{%
1}{G\left( t\right) }t^{s-1}dt\text{,} \label{Lfunc}
\end{equation}

\noindent defined for \textit{Re} $s > 1$ admits a holomorphic continuation to all complex values $s\neq 1$. For \textit{Re} $s>1 $ we define the Dirichlet series associated with the function $\widetilde{L}(s):=\sigma^{s-1} L\left( G,s\right) $ to be the series
\begin{equation}
\zeta_{G}(s)=\sum_{n=1}^{\infty }\frac{a_{n}}{n^{s}}.  \label{LG}
\end{equation}
\noindent Assuming that $G(t)\geq e^{\sigma t}-1$, the series $\zeta_G(s)$ is absolutely and
uniformly convergent for \textit{Re} $s>1$, and
\begin{equation}
\left\vert \sum_{n=1}^{\infty }\frac{a_{n}}{n^{s}}\right\vert \leq
\sum_{n=1}^{\infty }\frac{1}{n^{Res}} \text{.} \label{III}
\end{equation}
The proof of the statements (\ref{Lfunc})--(\ref{III}) can be found in \cite{Tempesta2}.

Clearly, each of the generalized zeta functions ($\ref{LG}$) is associated to a specific universality class, via the corresponding group exponential. Indeed, according to the construction discussed in the previous sections, assigning a group exponential is sufficient for determining completely a universality class.

Observe that the present construction is even more general, and can be proposed for logarithms different from that possessing the form (\ref{expLog}). In all these cases, under the above regularity hypothesis for $G(t)$, a L--function can be constructed, but its representation in terms of a series of the form ($\ref{LG}$) cannot be obtained in a simple way.

We observe that the class of uncorrelated systems does not possesses an associated zeta function: from this point of view, the absence of correlation is trivial. In particular, it would be nice to understand the statistical mechanical interpretation of the zeroes of the family of zeta functions described here.

\section{Central limit behaviour and other open problems}

In this work, we have introduced the notion of group entropy, which emerges naturally as a unifying tool for treating many theoretical aspects of generalized thermostatistics and provides a nontrivial connection with number theory.

A wide spectrum of theoretical perspectives can be explored. Here we suggest some open problems.

A key result in probability theory is the Central Limit Theorem (CLT). In its standard version, it states that the mean of a large number of independent random variables identically distributed, with finite expectation value and variance, has the normal distribution as an attractor \cite{Feller}. A central limit theorems for random correlated variables has been established \cite{VP}, \cite{HJU}. In the context of the present generalization of statistical mechanics, a natural question arises, which can be stated as follows.

Consider a set of random variables, with any p.d.f, correlated according to the law (\ref{correlation}). Establish whether, under suitable conditions, an analog of the CLT holds for the universality classes described by the group entropy (\ref{entropy}). Find the analytic expression of the attractor.
An interesting problem is to establish the exact relation between the present approach and the notion of "Gamma Entropy" recently proposed in \cite{AT}, i.e. under which conditions the universality classes described by Gamma entropies collapse into group entropies and vice versa.
In addition, it appears interesting to study the role of general correlations laws in complex networks (especially scale--free networks \cite{AB}), in biological complexity \cite{AOC}, in temporal series analysis and in ecology, in relation with the study of bio--diversity \cite{METAG}. Also, it would be important to deepen the comprehension of the dynamics of financial markets under nontrivial correlations (for a nonextensive approach see, e.g. \cite{Borland1}). An intriguing possibility is to classify economical systems according to the universality class of correlations that govern them.

\section*{Acknowledgments}

I am grateful to prof. C. Tsallis for a careful reading of the manuscript, encouragement and many useful discussions. I wish to thank prof. S. Thurner for stimulating discussions at the final stage of this work and in particular for drawing to my attention ref. \cite{AT}. I also thank prof. G. Mussardo for interesting discussions. Finally, I wish to thank the referees for useful comments.

I wish to thank the Centre of Mathematics for Applications (CMA), University of Oslo, Norway and the IMI--Madrid, Spain for granting me a \textit{"Abel Extraordinary Chair"}. Part of this work has been carried out during my stay at CMA.

The support from the Spanish Ministerio de Ciencia e Innovaci\'{o}n, research project FIS2008--00200 is gratefully acknowledged.

\end{document}